\documentclass[conference]{IEEEtran}

\usepackage{graphicx}                   
\usepackage{times,amsmath,amssymb,amsfonts}   
\usepackage{bm}                         
\usepackage{psfrag}                     
\usepackage{multirow}
\usepackage{cite}
\usepackage[table]{xcolor}
\usepackage{gensymb}
\usepackage{subcaption}
\usepackage{fancyhdr,lipsum}

\fancypagestyle{mahmood}{%
   \fancyhf{} 
   
   \fancyhead[C]{G. Alyami and I. N. Kostanic, \textquotedblleft On the spatial separation of multiuser channels using 73 GHz statistical channel models,\textquotedblright~accepted in IEEE International Conference on Communication, Networks and Satellite (COMNETSAT) (IEEE COMNETSAT 2016), Surabaya, Indonesia, Dec. 2016, pp. 14-18.}
}%

\makeatletter
\let\ps@IEEEtitlepagestyle\ps@mahmood
\makeatother

\definecolor{tui-green}{rgb}{0,0.455,0.478}
\definecolor{tui-blue}{rgb}{0,0.2,0.349}
\definecolor{tui-orange}{rgb}{1.0,0.475,0}
\definecolor{tui-lightblue}{rgb}{0.706,0.863,0.863}

\usepackage[load-configurations=abbreviations,load-configurations=binary]{siunitx}	
\DeclareSIUnit{\mms}{\milli\squaremetre}
\DeclareSIUnit{\inch}{in}
\DeclareSIUnit{\inchs}{in\squared}
\DeclareSIUnit{\mil}{mil}
\DeclareSIUnit{\Msps}{Msps}
\DeclareSIUnit{\Mbps}{Mbps}
\DeclareSIUnit{\LSB}{LSB}
\DeclareSIUnit{\pFS}{\percent FS}
\DeclareSIUnit{\dBc}{\deci\bel c}
\DeclareSIUnit{\dBm}{\deci\bel m}
\DeclareSIUnit{\dBFS}{\deci\bel FS}
\DeclareSIUnit{\dB}{\deci\bel}
\DeclareSIUnit{\dBi}{\deci\bel i}
\DeclareSIUnit{\hex}{0x}
\DeclareSIUnit{\vp}{\volt_{\text{p}}}
\DeclareSIUnit{\vpp}{\volt_{\text{pp}}}
\DeclareSIUnit{\kb}{\kilo\bit}
\DeclareSIUnit{\kB}{\kilo\byte}
\DeclareSIUnit{\MB}{\mega\byte}
\DeclareSIUnit{\GHz}{\giga\hertz}
\DeclareSIUnit{\MHz}{\mega\hertz}

\DeclareSIUnit{\mus}{\micro\second}
\DeclareSIUnit{\ns}{\nano\second}
\DeclareSIUnit{\fs}{\femto\second}

\title{On the Spatial Separation of Multi-user channels Using 73 GHz Statistical Channel Models}

\author{\IEEEauthorblockN{Geamel Alyami and Ivica Kostanic}
\IEEEauthorblockA{Wireless Center of Excellence (WiCE) \\
Florida Institute of Technology\\
Email: $\{$galyami2007$\}$@my.fit.edu} }

\begin{document}

\maketitle
\IEEEpeerreviewmaketitle
\begin{abstract}
This paper focuses at the investigation of the degree of orthogonality of channels of multiple users in densely populated indoor and outdoor scenarios. For this purpose, a statistical millimeter wave (mmwave) MIMO channel simulator is carefully designed using state of the art channel models. At the mmwave frequencies, human/vehicular mobility around the mobile users may partially or completely block the communication link. This give rise to the consideration of new channel modeling parameter i.e. probability of a user to be in LOS dynamics. Higher line of sight (LOS) probabilities may increase the spatial correlation among the multiuser channels. Therefore, quantification of the spatial separation of users in different scenarios with distinct LOS probabilities is crucial and it is the subject of investigation of this paper. Additionally, the mutual orthogonality of channels by changing the number of base station (BS) antennas and inter antenna element distance (IED) have also been investigated. Analysis shows that LOS blockage of certain scenario has more impact at the degree of orthogonality among users as compared to the number of BS antennas and the spacing between the antenna elements.
\end{abstract}
\vspace{3mm}
\section{Introduction}

An era to explore millimeter wave (mmwave) frequency bands (30-300 GHz) has now been started as they offer bandwidths of Ultra Wide Band (UWB) range and higher spatial resolution with antenna array gains. Millimeter wave frequencies are known to have quasi optical and specular propagation nature~\cite{802-11ad,Rappaport5G,RGB13}. This allows multiple users to be well separated in space and may greatly reduce the required signal processing efforts. At centimeter waves ($< 6$ GHz), it is almost understood that users can be spatially separated with base station antenna array having large number of antenna elements. Measurement results~\cite{FGD15} have provided an extensive analysis of spatial separation of users at 2.6 GHz band. Sub 6 GHz bands are rich in number of multipath components (MPC), and lower bandwidths allow wide sense stationary uncorrelated scattering assumption to hold~\cite{Mol09}. Therefore, intuitively at 2.6 GHz band with large directional antenna gains, multiple users can be well separated in space. On the other hand, mmWave channels are sparse due large channel bandwidths and the Rician K-factor is significantly high leading to high channel correlations~\cite{SamR15}. Therefore, LOS path component is assumed to be a dominant communication medium~\cite{MPK14}. Therefore at mmwave frequencies, if the LOS path is blocked/shadowed received signal power decreases significantly~\cite{Ref18}. From the BS perspectives such phenomena associated with mmwave frequencies may provide some advantages in terms of overall sum rate performance of the system in a densely populated scenario. For example, if more users are blocked then an apparently dense looking multiuser scenario will look like a sparse scenario at the BS which can now spatially separate different users. At mmwave 60 GHz frequency band, studies in~\cite{NHJ15} have investigated the spatial orthogonality of multiuser channels in an open square scenario. Degree of orthogonality between the users has been analyzed as function of inter antenna element distance (IED) and the distance between the users.  \\
The phenomena particularly associated with mmwave and light frequencies i.e. LOS blockage and its role in the spatial separation among the users needs a thorough investigation. With the best of our knowledge, such studies are yet missing and this paper is the first contribution investigating this effect. 
In the first part of the analysis, the multiuser orthogonality of indoor (shopping mall) and outdoor (open square) scenarios is compared as their LOS probability models differ significantly. Finally, the role of different number of antenna elements and their IED have also been analyzed.\\
Rest of the paper is organized as follows:~Section.~\ref{sec:Sysmod} extensively describes the experiment scenario and the channel models that have been used to derive the statistical MIMO channel simulator. Section.~\ref{sec:Simulation} provides an in-depth analysis of different channel model parameters that play a role in defining the correlation or spatial separation of the users in the cell. Finally, all the findings are concluded in~Section.~\ref{sec:CONCLUSION} with a short description of the future work.   

\section{Channel Model}
\label{sec:Sysmod}
\begin{figure}[t!]
\begin{center}
\includegraphics[width=1.00\columnwidth]{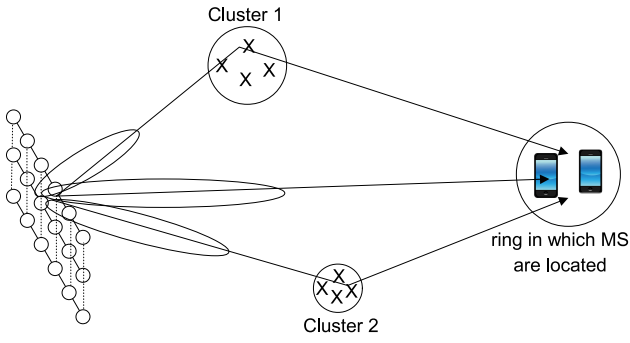}
\end{center}
\caption{Channel model and experiment description}
\label{scenario}
\end{figure}

\begin{figure}[!t]
    \centering
    \begin{subfigure}[b]{0.235\textwidth}
   	    \includegraphics[width=0.95\columnwidth]{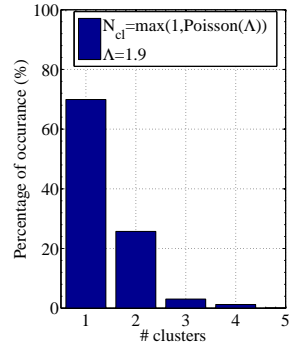}
        \caption{Model for $N_{cl}$}
        \label{Ncl_model}
    \end{subfigure}
    ~ 
    \begin{subfigure}[b]{0.235\textwidth}
   	    \includegraphics[width=0.95\columnwidth]{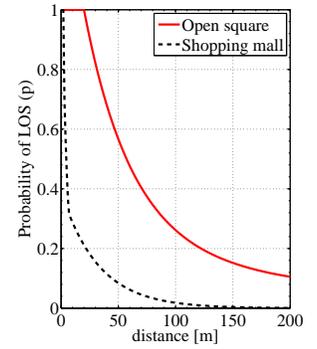}
        \caption{Probability of LOS}
        \label{plos}
    \end{subfigure}
    \caption{(a) Number of clusters and the (b) probability of LOS models of the considered propagations scenarios}\label{fig:models}
\end{figure}

Consider a downlink multiuser MIMO system with $n_\text{BS}$ antennas mounted at the base station (BS) of height $h_\text{BS}$ and $n_\text{U}$ mobile users (MU) each of height $h_\text{U}$. All the users are closely located in the XY plane of the Cartesian coordinate system inside a ring of radius $R$ as shown in~Fig.~\ref{scenario}. Note that the heights of the BS and MU correspond to the z-axis of the coordinate system. Inside the ring, location of the users and their antenna orientations changes following a uniform random distribution in each channel snapshot. All users are assumed to be static for a particular channel realization, hence the channel stays time invariant. Let $ N_{cl} $ be the number of single bounce multipath clusters proposed in \cite{Ref7RapaportmmCapacity} for the considered 73 GHz frequency band. 
\begin{equation}
\label{ncl}
N_{cl}=\max\left(1,\text{Poisson}\left(\Delta \right)  \right), \quad \Delta=0.9
\end{equation}
Fig.~\ref{Ncl_model} shows the percentage of number of multipath clusters generated as a result of the model in~Eq.~\ref{ncl}. It shows that the probability of 1-2 cluster dominates in propagation scenario. Note that each one of the multipath clusters is composed of $ N_{ray} $ subpaths. Let $\theta_{i,l}^\text{A} $ and $\theta_{i,l}^\text{D}$ be azimuth angle of arrival and departure respectively and they belong to the $l^\text{th}$ ray of the $i^\text{th}$ cluster. Similarly $\phi_{i,l}^\text{A} $ and $\phi_{i,l}^\text{D}$ correspond to the elevation angles of arrival and departure respectively. Let $K$ be the number of antennas at each MU, then the channel matrix $ \bm{H}_i(\tau )\in \mathbb{C}^{n_\text{BS}\times K} $ of a single user MIMO link can be written as 
\begin{equation}
\label{HLTI}
\begin{array} {lcl}
\bm{H}_i\left( \tau\right) & = &  \sum_{i=1}^{N_{cl}} \sum_{l=1}^{N_{ray}} \alpha_{i,l} \sqrt{L\left( r_{i,l}\right) } \bm{a}_r\left( \phi_{i,l}^A , \theta_{i,l}^A\right) \times \\ &  &  \bm{a}_t^H\left( \phi_{i,l}^D,\theta_{i,l}^D\right) h\left( \tau -\tau_{i,l}\right)  + \bm{H}_\text{LOS}\left( \tau\right) 
\end{array}
\end{equation}
where,
\begin{itemize}
\item $\bm{H}_\text{LOS}\left( \tau\right)$ is the channel impulse response matrix of the LOS component arriving at delay $\tau$.
\item $\bm{a}_t$ and $\bm{a}_r$ are the transmit and receive antenna array steering vectors.
\item $h\left( \tau -\tau_{i,l}\right)$ is the impulse response at the delay tap $\tau_{i,l}$ of the $l^\text{th}$ ray of the $i^\text{th}$ cluster.
\item $\alpha_{i,l}$ and $L\left( r_{i,l}\right)$ are the complex channel gain and attenuation loss of the $l^\text{th}$ ray of the $i^\text{th}$ cluster respectively.
\end{itemize}
Since each sub-path may have different scattering conditions so the sub-path phases can be considered as independent identical distribution. To simplify the model, it has been assumed that there exists a constant distance between the elements of antenna array and all the scatterers belonging to the same cluster. Let $r_{i}$ be the propagation distance between BS and MU via the central ray of the $i^\text{th}$ cluster. The propagation distance of subpaths associated to the $i^\text{th}$ cluster is geometrically computed as 
\begin{equation}
\label{ril}
\begin{array} {lcl}
r_{i,l} =  r_i+ \\  \sqrt{\left( h_\text{BS}-h_\text{U}+ r_i \sin\theta_{i,l}^D\right) ^2+\left( d-r_i\cos\theta_{i,l}^D\cos\phi_{i,l}^D\right) ^2}
\end{array}\cdot
\end{equation}
For the attenuation of the link, the channel simulator uses the same model proposed in~\cite{Ref18}
\begin{equation} \label{Lril}
\begin{split}
L\left ( r_{i,l} \right ) & = -20\log_{10}\left(\frac{4\pi}{\lambda} \right) \\
 & - 10n\left [ 1-b+\frac{bc}{\lambda f_{0}} \right ]\log_{10}\left ( r_{i,l} \right )-X_{\sigma } 
\end{split}
\end{equation}
where $\lambda$ is the wavelength of the center frequency, $n$ is the path loss exponent, $b$ is a system parameter, $f_0$ is the fixed reference frequency (the centroid of all frequencies represented by the path loss model). Finally, $X_{\sigma }$ is the shadow fading term with zero mean and variance $\sigma^2$. Parameters values of~Eq.~\ref{Lril} for different propagation scenarios have been defined in~\cite{Ref18}. \\
Millimeter wave frequencies are known for their quasi optical nature~\cite{Malt10}, and similar like light multipath components could be blocked. Blockage modeling of LOS component is of fundamental interest as it caries the highest energy. Therefore, channel model for LOS component is defined as  
\begin{equation}
\label{HLOSI}
\begin{array} {lcl}
\bm{H}_\text{LOS}\left(\tau \right) & = & I_\text{LOS}\left ( d \right )\sqrt{Kn_\text{BS}}e^{j\eta }\sqrt{L\left ( d \right )} \times \\ &  &  \bm{a}_r\left( \phi_{i,l}^A , \theta_{i,l}^A\right)\bm{a}_t^H\left( \phi_{i,l}^D,\theta_{i,l}^D\right) h\left( \tau -\tau_{i,l}\right)   
\end{array}
\end{equation}
where, $\eta\sim \mathcal{U}\left(0,2\pi \right) $ corresponds to the phase of the LOS component and  $I_\text{LOS}\left(d \right) $ is a random variable indicating if a LOS link is present between the transmitter and receiver at a distance $d$. Regarding the probability $ p $  of a LOS link, the same models defined in~\cite{Ref18,Ref19} are used for the evaluation of an open square and shopping mall scenarios as shown in~Eq.~\ref{pOnH} and~Eq.~\ref{pInH} respectively. 
\begin{equation}
\label{pOnH}
p_\text{open square}=\min\left ( \frac{20}{d} \right )\left ( 1-e^{-\frac{d}{39}} \right )+e^{-\frac{d}{39}}
\end{equation}
\begin{equation}
\label{pInH}
p_\text{shopping mall}=\left\{\begin{matrix}
& 1 \quad \quad \quad \quad \quad \quad \quad  d \leq 1.2  \\ 
& e^{-(\frac{d-1.2}{4.7})} \quad   1.2< d \leq 6.5 \\
& 0.32e^{-(\frac{d-1.2}{4.7})} \quad \quad  d\geq 6.5
\end{matrix}\right.
\end{equation}
Fig.~\ref{plos} shows the results of the models defined in~Eq.~\ref{pOnH} and~Eq.~\ref{pInH} and it can be seen that LOS probability of users is significantly higher in open square environment as compared to shopping mall scenario. Some other selected statistical features and model distributions are summarized in~Table.~\ref{statTab} whereby other detailed descriptions are available in~\cite{BuzziD16}. Multiuser MIMO channel matrix is now defined as 
\begin{equation}
\label{HLTI2}
\bm{H}=\left[  \bm{H}_1\quad \bm{H}_2\cdots\bm{H}_{n_\text{U}} \right] \in  \mathbb{C}^{n_\text{BS}\times n_\text{U}K} 
\end{equation}
Let $\sigma_{1}>\sigma_{2}>...>\sigma_{n_\text{U}K}$ be the singular values of $\bm{H}$ then we define
\begin{equation}
\label{cond}
\kappa\left(\bm{H} \right) =\frac{\sigma_{\max}}{\sigma_{\min}}
\end{equation}
If the channel is ill-conditioned or highly correlated $\kappa\left(\bm{H} \right)$ would be very large. In this paper however, we used $\frac{1}{\kappa\left(\bm{H} \right)}=\frac{\sigma_{\min}}{\sigma_{\max}}$ where $0\leq\frac{\sigma_{\min}}{\sigma_{\max}}\leq 1$ to analyze the singular value spread of the channel which in other words define the degree of orthogonality among the users.  The ratio $  \frac{\sigma_{\min}}{\sigma_{\max}} \rightarrow 0 $ shows that the channels between users are fully correlated and it is difficult to spatially separate the users. On the other hand, $  \frac{\sigma_{\min}}{\sigma_{\max}} \rightarrow 1 $ indicates that the degree of orthogonality is high among users.  
\begin{table}[!t]
\centering
\caption{Table of statistical distributions along with some other generic model parameters}
\label{statTab}
\begin{tabular}{|l|l|l|l|l}
\cline{1-4}
\cellcolor[HTML]{FFCC67}Parameter                                                    & \cellcolor[HTML]{FFCC67}Distribution  & \cellcolor[HTML]{FFCC67}Mean                                 & \cellcolor[HTML]{FFCC67}\begin{tabular}[c]{@{}l@{}}St. \\ dev.\end{tabular} &  \\ \cline{1-4}
Azimuth AOA per cluster $i$                                                          & Laplacian                             & $ \theta_i^A \sim \mathcal{U}[0,2\pi] $                      & $ 5^{\circ}$                                                               &  \\ \cline{1-4}
Azimuth AOD per cluster $i$                                                          & Laplacian                             & $\theta_i^D\sim \mathcal{U}[\frac{-\pi}{2},\frac{\pi}{2}] $  & $ 5^{\circ}$                                                               &  \\ \cline{1-4}
Elevation AOA per cluster $i$                                                        & Laplacian                             & $\phi_i^A\sim \mathcal{U} [\frac{-\pi}{2},\frac{\pi}{2}] $   & $ 5^{\circ}$                                                               &  \\ \cline{1-4}
Elevation AOD per cluster $i$                                                        & Laplacian                             & $\phi_i^D\sim \mathcal{U} [\frac{-\pi}{2},\frac{\pi}{2}]$    & $ 5^{\circ}$                                                               &  \\ \cline{1-4}
\# Clusters$\left(N_{cl}\right) $                                                   & \multicolumn{3}{l|}{$ \max \left\lbrace \text{Poisson}(\Lambda),1\right\rbrace,\Lambda=1.9$\cite{Ref7RapaportmmCapacity}}                                                               &  \\ \cline{1-4}
\# Rays per cluster $i$                                                              & \multicolumn{3}{l|}{$\mathcal{U} [1,30]$ \cite{Ref14GlobcomRapaport}}                                                                                                            &  \\ \cline{1-4}
\multicolumn{4}{|c|}{\cellcolor[HTML]{EFEFEF}\textbf{\underline{Misc. parameters}}}                                                                                                                                                                                                            &  \\ \cline{1-4}
\cellcolor[HTML]{FFCC67}Parameter                                                    & \multicolumn{3}{l|}{\cellcolor[HTML]{FFCC67}Value}                                                                                                                                 &  \\ \cline{1-4}
Scenario                                                                             & \multicolumn{3}{l|}{\begin{tabular}[c]{@{}l@{}}Indoor-Shopping mall\\ Outdoor- Open square\\  \end{tabular}}                                                                                                                                                              &  \\ \cline{1-4}
Carrier frequency                                                                    & \multicolumn{3}{l|}{73 GHz}                                                                                                                                                              &  \\ \cline{1-4}
$ n_\text{BS} $                                                                      & \multicolumn{3}{l|}{\begin{tabular}[c]{@{}l@{}}$5\times8 $ UPA $\quad\rightarrow 40$\\ $10\times8 $ UPA$\quad\rightarrow 80$\\ $15\times8 $ UPA$\quad\rightarrow 120$\\ $20\times8 $ UPA$\quad\rightarrow 160$\end{tabular}} &  \\ \cline{1-4}
$ n_\text{U} $                                                                       & \multicolumn{3}{l|}{2,5,10}                                                                                                                                                              &  \\ \cline{1-4}
Antenna spacing $ d_\text{a} $                                                 &  \multicolumn{3}{l|}{$0.5 \lambda, 4 \lambda,6 \lambda$}                                                                                                                                                               &  \\ \cline{1-4}
\# antennas per user                                                                 &  \multicolumn{3}{l|}{1}                                                                                                                                                              &  \\ \cline{1-4}
$ h_\text{BS} $                                                                      &  \multicolumn{3}{l|}{ 7m}                                                                                                                                                              &  \\ \cline{1-4}
$ h_\text{MU} $                                                                      &   \multicolumn{3}{l|}{1.68 m}                                                                                                                                                              &  \\ \cline{1-4}
\begin{tabular}[c]{@{}l@{}}Average BS-MU \\ distance $\left( d\right) $\end{tabular} &   \multicolumn{3}{l|}{20 m}                                                                                                                                                              &  \\ \cline{1-4}
Radius of the ring $\left( R\right) $                                                &  \multicolumn{3}{l|}{5 m}                                                                                                                                                              &  \\ \cline{1-4}
BS orientation                                                                       &  \multicolumn{3}{l|}{Arbitrary}                                                                                                                                                              &  \\ \cline{1-4}
MS orientation                                                                       &  \multicolumn{3}{l|}{Arbitrary}                                                                                                                                                              &  \\ \cline{1-4}
\end{tabular}
\end{table}

\section{Simulation Results}
\label{sec:Simulation} 

\begin{figure}[t!]
\begin{center}
\includegraphics[width=0.95\columnwidth]{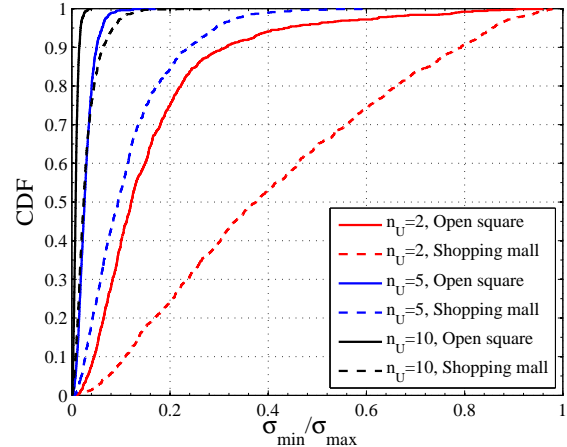}
\end{center}
\caption{CDF of $ \frac{\sigma_{\min}}{\sigma_{\max}} $ for two scenarios, $20\times8 $ UPA at BS, $  d_a=0.5\lambda $ }
\label{Scen_Comp}
\end{figure}

\begin{figure}[t!]
\begin{center}
\includegraphics[width=0.95\columnwidth]{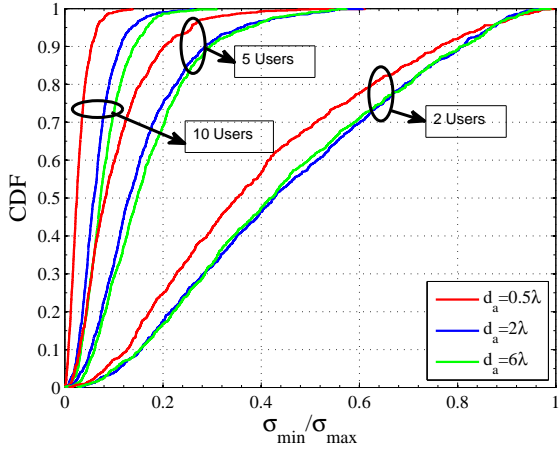}
\end{center}
\caption{CDF of $ \frac{\sigma_{\min}}{\sigma_{\max}} $ by changing $ d_a $, $5\times8 $ UPA at BS, $ N_{cl} $-- Model based, shopping mall scenario }
\label{IED}
\end{figure}

For the simulations, $n_\text{U}$ users are located inside a ring of 5 meters radius in the XY plane of the Cartesian coordinate system at a distance 20 meters away from BS. It is assumed that $ N_{cl} $ clusters are available in propagation environment including LOS path and none of the clusters is shared by two or more users. Detailed simulation parameters are defined in Table.~\ref{statTab}.\\
As discussed earlier that mmwave frequencies are quasi optical in nature and they could be shadowed or completely blocked by the many environmental factors such as human or vehicular mobility around the MU. Therefore, it is much likely that under certain propagation setup, some but not all users in the our experimental setup (see~Fig.~\ref{scenario}) may be in pseudo LOS dynamics or in worst case they are completely blocked. It is shown in~Fig.~\ref{plos} that the in open square scenario, probability of LOS is significantly higher than the shopping mall scenario. In other words, it means that in shopping mall scenario more users could be blocked or shadowed. Hence, from BS perspective, a dense looking shopping mall scenario does not remain so dense due higher users blockage probabilities. As a proof of concept,~Fig.~\ref{Scen_Comp} shows that singular values spread $\frac{\sigma_{\min}}{\sigma_{\max}}$ is lower in open square as compared to the shopping mall scenario for all the considered number of users inside the ring $R$. It clearly shows that users are more correlated. On the other hand, users in the shopping mall scenario seems to be low correlated as more users are blocked.\\ 
Increase in the IED among the antennas reduces the spatial correlation of multipath components. Dual polarized antenna arrays are also an effective way to reduce to the spatial correlation~\cite{WNTC15}. However, in this paper the analysis is limited only to IED and dual polarization effects are left as future extensions of the subjected topic. Results in Fig.~\ref{IED} discuss the case when $N_{cl}$ is still low and it is generated by the model shown in~Table.~\ref{statTab}. It has been observed that increased IED considerably reduces the singular values spread, however at $ 6 \lambda $, the results are bottlenecked and don't improve the channel conditions. This behavior is almost consistent for 3 cases of number of users inside the ring. Therefore, it can be argued that increased IED reduces the antenna correlation at the BS side but the inter-user spatial correlation still prevails in the channel. \\
\begin{figure}[t!]
\begin{center}
\includegraphics[width=0.95\columnwidth]{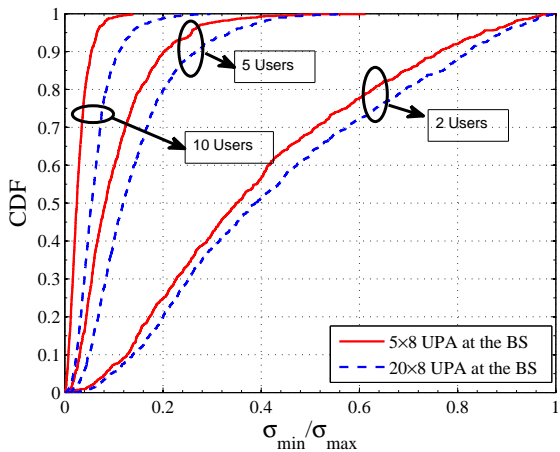}
\end{center}
\caption{CDF of $ \frac{\sigma_{\min}}{\sigma_{\max}} $ by changing BS antennas, $ N_{cl} $- Model based, $  d_a=0.5\lambda $, shopping mall scenario}
\label{AntennaEffect}
\end{figure}
Generally, it is assumed that the large number of antenna elements in the array result in phenomena known as \textit{channel hardening}~\cite{NC14}. In mathematical sense it means that the off-diagonal elements of the covariance matrix $\bm{H}^H\bm{H}$ are either close to zero or more optimistically they are exactly zeros. A benefit of \textit{channel hardening} is that signal processing efforts become negligibly small as the low complexity match filtering precoder $\bm{W}=\bm{H}^H$ becomes nearly optimal. However, channel hardening also requires a rich scattering environment where multipath components arrive from all spatial directions. An example of such propagation environment is the Jakes channel model~\cite{Jakes:1994}. However, in mmwave frequencies multipath components come from very few spatial directions a.k.a spatial lobes\cite{SamimiR15}.\\
In Fig.~\ref{AntennaEffect}, the antenna spacing is set to a fixed value i.e. $d_a=0.5\lambda$ and the multi-path cluster model of \cite{Ref7RapaportmmCapacity} is used to evaluate the impact of large number antenna elements in the array. The results show that increasing the number of antennas in UPA of BS from $n_\text{BS}=5\times8 $ UPA to $n_\text{BS}=20\times8 $ UPA does not give a significant improvement in the channel condition number as compared to reducing the number of users in the ring. This is understandable as number of time clusters are lower than 4 where 1 cluster has the highest probability as shown in~Fig.~\ref{Ncl_model}. These time cluster come from very few number of spatial directions and with small azimuth angular spread per cluster i.e. $5^{\circ}$ it is hard to spatially separate the users in the small ring. This clearly indicates that unlike massive MIMO systems at centimeter waves, increasing only the BS antennas at mmWave may not be sufficient to spatially separate the closely located users.

\section{Conclusion And Future Work}
\label{sec:CONCLUSION}
In this paper, spatial separation of users in densely populated shopping mall and open square scenarios have been analyzed for 73 GHz millimeter frequency band. Investigation considers multiple propagation features and models of mmwave frequencies to analyze the extent upto which different number of users can be spatial separated. Results show that from sum rate perspective higher blockage probability is not always a negative feature of mmwave channels. In a densely populated scenarios, it may have positive impact as user separation becomes easier if more users are being blocked. Additionally, the number of antenna elements in the BS array and IED also play a role in increasing the orthogonality between the channels of multi-user MIMO systems. However, the impact of LOS probability on user separation is much more significant as compared to IED and number of antenna elements in the BS array.

\bibliographystyle{IEEEtran}

\begin{thebibliography}{10}
\providecommand{\url}[1]{#1}
\csname url@samestyle\endcsname
\providecommand{\newblock}{\relax}
\providecommand{\bibinfo}[2]{#2}
\providecommand{\BIBentrySTDinterwordspacing}{\spaceskip=0pt\relax}
\providecommand{\BIBentryALTinterwordstretchfactor}{4}
\providecommand{\BIBentryALTinterwordspacing}{\spaceskip=\fontdimen2\font plus
\BIBentryALTinterwordstretchfactor\fontdimen3\font minus
  \fontdimen4\font\relax}
\providecommand{\BIBforeignlanguage}[2]{{%
\expandafter\ifx\csname l@#1\endcsname\relax
\typeout{** WARNING: IEEEtran.bst: No hyphenation pattern has been}%
\typeout{** loaded for the language `#1'. Using the pattern for}%
\typeout{** the default language instead.}%
\else
\language=\csname l@#1\endcsname
\fi
#2}}
\providecommand{\BIBdecl}{\relax}
\BIBdecl

\bibitem{802-11ad}
\BIBentryALTinterwordspacing
A.~Maltsev, V.~Erceg, E.~Perahia, C.~Hansen, R.~Maslennikov, A.~Lomayev,
  A.~Sevastyanov, G.~Morozov, M.~Jacob, S.~Priebe, T.~Kürner, S.~Kato,
  H.~Sawada, K.~Sato, and H.~Harada, ``Channel models for 60 ghz wlan
  systems,'' Tech. Rep. IEEE 802.11-09/334r3, July 2009. [Online]. Available:
  \url{https://mentor.ieee.
  org/802.11/dcn/09/11-09-0334-03-00ad-channel-models-for-60-ghz-wlan-systems.doc}
\BIBentrySTDinterwordspacing

\bibitem{Rappaport5G}
T.~Rappaport, S.~Sun, R.~Mayzus, H.~Zhao, Y.~Azar, K.~Wang, G.~Wong, J.~Schulz,
  M.~Samimi, and F.~Gutierrez, ``Millimeter wave mobile communications for 5g
  cellular: It will work!'' \emph{Access, IEEE}, vol.~1, pp. 335--349, 2013.

\bibitem{RGB13}
T.~S. Rappaport, F.~Gutierrez, E.~Ben-Dor, J.~N. Murdock, Y.~Qiao, and J.~I.
  Tamir, ``Broadband millimeter-wave propagation measurements and models using
  adaptive-beam antennas for outdoor urban cellular communications,''
  \emph{IEEE Transactions on Antennas and Propagation}, vol.~61, no.~4, pp.
  1850--1859, April 2013.

\bibitem{FGD15}
J.~Flordelis, X.~Gao, G.~Dahman, F.~Rusek, O.~Edfors, and F.~Tufvesson,
  ``Spatial separation of closely-spaced users in measured massive multi-user
  mimo channels,'' in \emph{IEEE International Conference on Communications
  (ICC)}, June 2015, pp. 1441--1446.

\bibitem{Mol09}
A.~F. Molisch, ``Ultra-wide-band propagation channels,'' \emph{Proceedings of
  the IEEE}, vol.~97, no.~2, pp. 353--371, Feb 2009.

\bibitem{SamR15}
\BIBentryALTinterwordspacing
M.~K. Samimi and T.~S. Rappaport, ``28 ghz millimeter-wave ultrawideband
  small-scale fading models in wireless channels,'' \emph{CoRR}, vol.
  abs/1511.06938, 2015. [Online]. Available:
  \url{http://arxiv.org/abs/1511.06938}
\BIBentrySTDinterwordspacing

\bibitem{MPK14}
A.~Maltsev, A.~Pudeyev, I.~Karls, I.~Bolotin, G.~Morozov, R.~Weiler, M.~Peter,
  and W.~Keusgen, ``Quasi-deterministic approach to mmwave channel modeling in
  a non-stationary environment,'' in \emph{2014 IEEE Globecom Workshops (GC
  Wkshps)}, Dec 2014, pp. 966--971.

\bibitem{Ref18}
\emph{{"5G Channel Model for bands up to 100 GHz"}},
  http://www.5gworkshops.com/5GCM.html/.

\bibitem{NHJ15}
S.~L.~H. Nguyen, K.~Haneda, J.~Jarvelainen, A.~Karttunen, and J.~Putkonen, ``On
  the mutual orthogonality of millimeter-wave massive mimo channels,'' in
  \emph{IEEE 81st Vehicular Technology Conference (VTC Spring)}, May 2015, pp.
  1--5.

\bibitem{Ref7RapaportmmCapacity}
M.~R. Akdeniz, Y.~Liu, M.~K. Samimi, S.~Sun, S.~Rangan, T.~S. Rappaport, and
  E.~Erkip, ``Millimeter wave channel modeling and cellular capacity
  evaluation,'' \emph{IEEE Journal on Selected Areas in Communications},
  vol.~32, no.~6, pp. 1164--1179, June 2014.

\bibitem{Malt10}
A.~Maltsev, R.~Maslennikov, A.~Sevastyanov, A.~Lomayev, and A.~Khoryaev,
  ``Statistical channel model for 60 ghz wlan systems in conference room
  environment,'' in \emph{Proceedings of the Fourth European Conference on
  Antennas and Propagation}, April 2010, pp. 1--5.

\bibitem{Ref19}
\BIBentryALTinterwordspacing
K.~Haneda, L.~Tian, Y.~Zheng, H.~Asplund, J.~Li, Y.~Wang, D.~Steer, C.~Li,
  T.~Balercia, S.~Lee, Y.~Kim, A.~Ghosh, T.~A. Thomas, T.~Nakamura,
  Y.~Kakishima, T.~Imai, H.~C. Papadopoulos, T.~S. Rappaport, G.~R.~M. Jr.,
  M.~K. Samimi, S.~Sun, O.~H. Koymen, S.~Hur, J.~Park, J.~C. Zhang, E.~Mellios,
  A.~F. Molisch, S.~S. Ghassamzadah, and A.~Ghosh, ``5g 3gpp-like channel
  models for outdoor urban microcellular and macrocellular environments,''
  \emph{CoRR}, vol. abs/1602.07533, 2016. [Online]. Available:
  \url{http://arxiv.org/abs/1602.07533}
\BIBentrySTDinterwordspacing

\bibitem{BuzziD16}
\BIBentryALTinterwordspacing
S.~Buzzi and C.~D'Andrea, ``On clustered statistical mimo millimeter wave
  channel simulation,'' \emph{CoRR}, vol. abs/1604.00648, 2016. [Online].
  Available: \url{http://arxiv.org/abs/1604.00648}
\BIBentrySTDinterwordspacing

\bibitem{Ref14GlobcomRapaport}
M.~K. Samimi and T.~S. Rappaport, ``Statistical channel model with
  multi-frequency and arbitrary antenna beamwidth for millimeter-wave outdoor
  communications,'' in \emph{2015 IEEE Globecom Workshops (GC Wkshps)}, Dec
  2015, pp. 1--7.

\bibitem{WNTC15}
W.~Ahmad, N.~Iqbal, T.~A. Shahzad, and C.~Schneider, ``Pairwise correlation and
  performance analysis of mimo systems in indoor scenarios,'' in \emph{IEEE
  International Conference on Aerospace Electronics and Remote Sensing
  Technology (ICARES)}, Dec 2015, pp. 1--6.

\bibitem{NC14}
T.~L. Narasimhan and A.~Chockalingam, ``Channel hardening-exploiting message
  passing (chemp) receiver in large-scale mimo systems,'' \emph{IEEE Journal of
  Selected Topics in Signal Processing}, vol.~8, no.~5, pp. 847--860, Oct 2014.

\bibitem{Jakes:1994}
W.~C. Jakes and D.~C. Cox, Eds., \emph{Microwave Mobile Communications}.\hskip
  1em plus 0.5em minus 0.4em\relax Wiley-IEEE Press, 1994.

\bibitem{SamimiR15}
\BIBentryALTinterwordspacing
M.~K. Samimi and T.~S. Rappaport, ``3-d statistical channel model for
  millimeter-wave outdoor mobile broadband communications,'' \emph{CoRR}, vol.
  abs/1503.05619, 2015. [Online]. Available:
  \url{http://arxiv.org/abs/1503.05619}
\BIBentrySTDinterwordspacing

\end{thebibliography}

\end{document}